\documentclass{svproc}
\usepackage{url}
\usepackage{tabularx}
\usepackage{graphicx}

\usepackage{subcaption}
\usepackage{hyperref}

\begin{document}
\mainmatter              
\title{Convolutional neural networks for automatic detection of  Focal Cortical Dysplasia}
\titlerunning{CNNs for FCD Detection}  
%
\author{Ruslan Aliev\inst{1} \and Ekaterina Kondrateva \inst{1} \and Maxim Sharaev\inst{1} \and Oleg Bronov\inst{2} \and Alexey Marinets\inst{2} \and Sergey Subbotin\inst{2} \and Alexander Bernstein\inst{1} \and Evgeny Burnaev\inst{1} }
\authorrunning{R. Aliev et al.} 
%
%
\institute{Skolkovo Institute of Science and Technology, Moscow, Russia,\\
\email{ruslan.aliev@skoltech.ru} 
\and  National Medical and Surgical Center named after N.I. Pirogov, Moscow, Russia \\
\email{}}

\maketitle              

\begin{abstract}
Focal cortical dysplasia (FCD) is one of the most common epileptogenic lesions associated with cortical development malformations. However, the accurate detection of the FCD relies on the radiologist professionalism, and in many cases, the lesion could be missed. In this work, we solve the problem of automatic identification of FCD on magnetic resonance images (MRI). For this task, we improve recent methods of Deep Learning-based FCD detection and apply it for a dataset of 15 labeled FCD patients. The model results in the successful detection of FCD on 11 out of 15 subjects.

\end{abstract}
\section{Introduction}
The imaging and computational technologies' latest achievements facilitated the rapid development of the decision support systems based on medical images \cite{shin2016deep}. Application of the machine learning and computer vision methods in medical imagery showed that the prediction and prognosis could be made automatically to identify pathological changes and make a medical prognosis \cite{sharaev2018mri}. At the same time, the high cost of the manual radiological examination determines the urge to develop such systems for the automated analysis in neuroimaging data as well as using machine learning methods to support the radiological process, increase the effectiveness and quality of the diagnostics.

Focal cortical dysplasia (FCD) is one of the most common epileptogenic lesions associated with malformations of cortical development \cite{tassi2002focal}.  The development of a decision support system for radiologists in FCD detection is important, and it urges large and rigorously annotated datasets. However, even on small samples with a weak annotation, the FCD detection could be enhanced with the means of deep learning tricks, such as autoencoder pretraining on unannotated data. 

We acknowledge our project medical partner science-practical center n.a. Pirogov for annotated data. We acknowledge Zhores HPC cluster for the computation facilities. All code for experiments is publicly available\footnote{\href{https://github.com/alievrusik/cnn\_fcd\_detection}{\texttt{https://github.com/alievrusik/cnn\_fcd\_detection}}}.

\subsection{Related works}
Currently, there are plenty of methods for automatic FCD detections. Previous approaches rely on the brain's symmetric structure, template matching, and feature estimations, using conventional machine learning techniques. Some statistics (e.g., cortical thickness, GM-WM blur, T1 hyper-intensity) are estimated and fed into decision system (e.g. Support Vector Machine) \cite{besson2008surface}, \cite{colliot2006segmentation}. Such methods require severe feature engineering.

Recently, deep-learning-based methods were proposed for FCD detection. In \cite{dev2019automatic} authors use U-Net like architecture for building FCD segmentation model. They train the network using sagittal slices and get DICE score of 52.47, which is superior to non-neural-networks approaches. In \cite{wang2020automated}, authors combine classical computer vision and deep learning techniques using special patch extraction techniques, CNN classification, and post-processing. It yields state-of-the-art results in terms of accuracy (classification) and Intersection over Union (detection).

In this paper, we reproduce the state-of-the-art method proposed in \cite{wang2020automated}, and made several significant improvements to enhance its performance on the dataset provided by our medical partners. 

\section{Data and methods}
 For training we used 15 labeled FCD subjects (Pirogov), 15 unlabeled FCD subjects (Pirogov) and 17 healthy control subjects from HCP \cite{van2013wu} and LA5 \cite{poldrack2013toward} datasets. The subjects were scanned by the SIEMENS MAGNETRON Scyra 3T scanner with resolution 256x256x256 voxels. In all experiments, we use leave-one-out (LOO) validation scheme.
 
\begin{figure}[ht!]
  \centering
  \begin{subfigure}{\linewidth}
      \includegraphics[width=\linewidth]{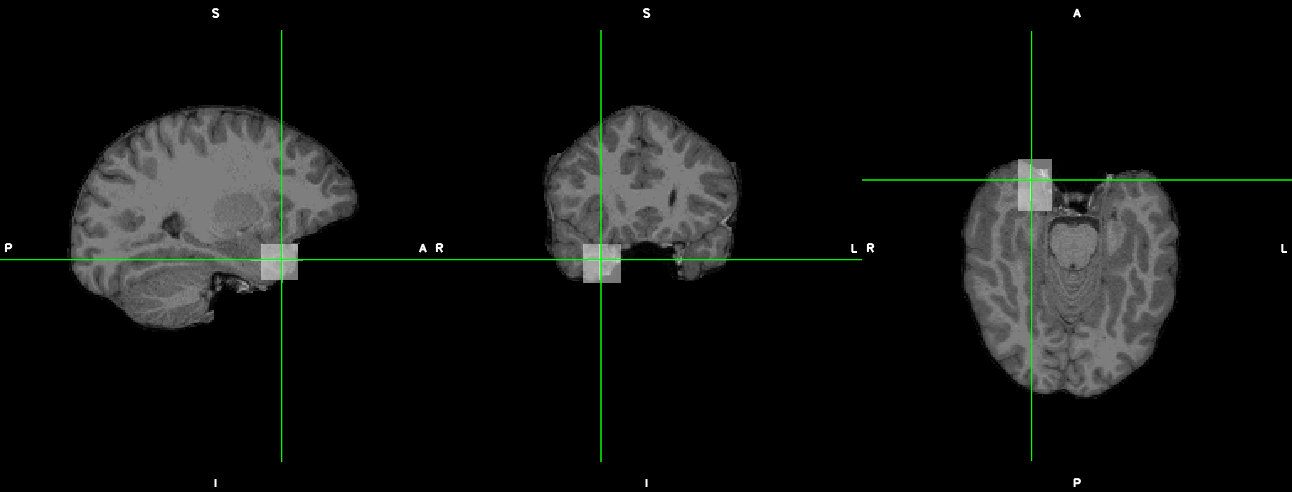}
    \caption{2D rectangular boxes provided by professional radiologist for each view.}  
      \label{fig:ex_of_vis} 

  \end{subfigure}
  \begin{subfigure}{.45\linewidth}
       \includegraphics[width=\linewidth]{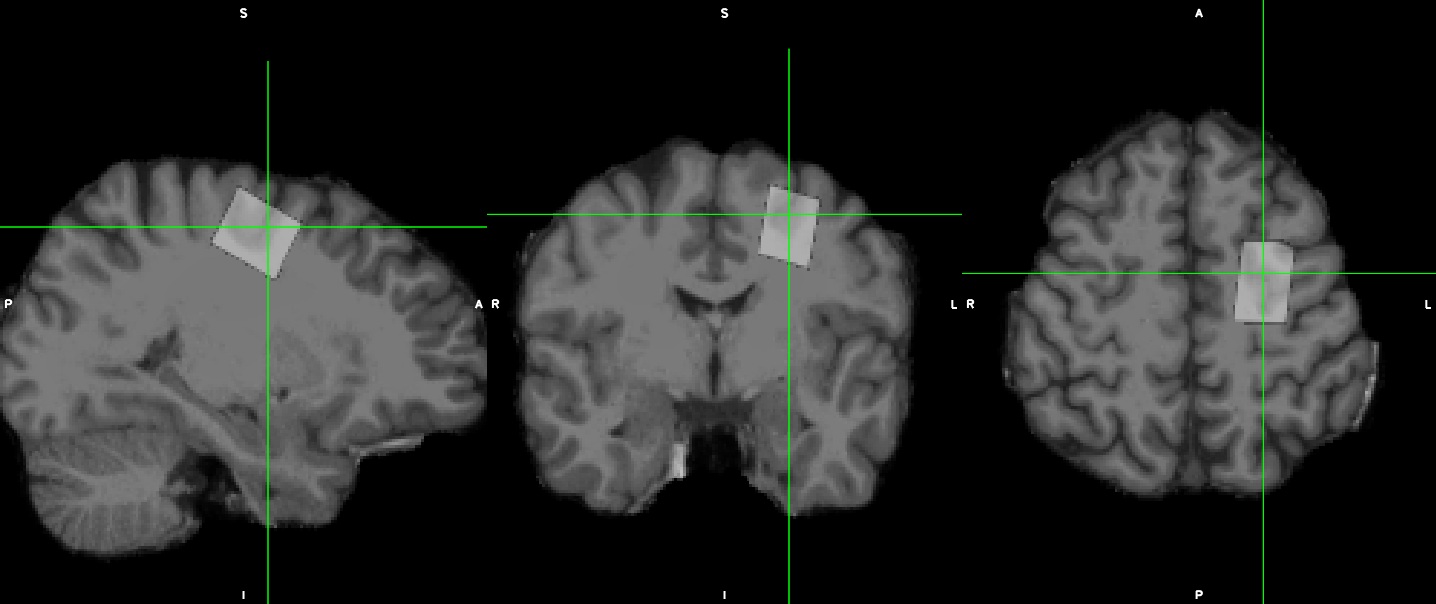}
  \caption{Rectangular mask (right).}
  \label{fig:rect} 
  \end{subfigure}
  \begin{subfigure}{.45\linewidth}
       \includegraphics[width=\linewidth]{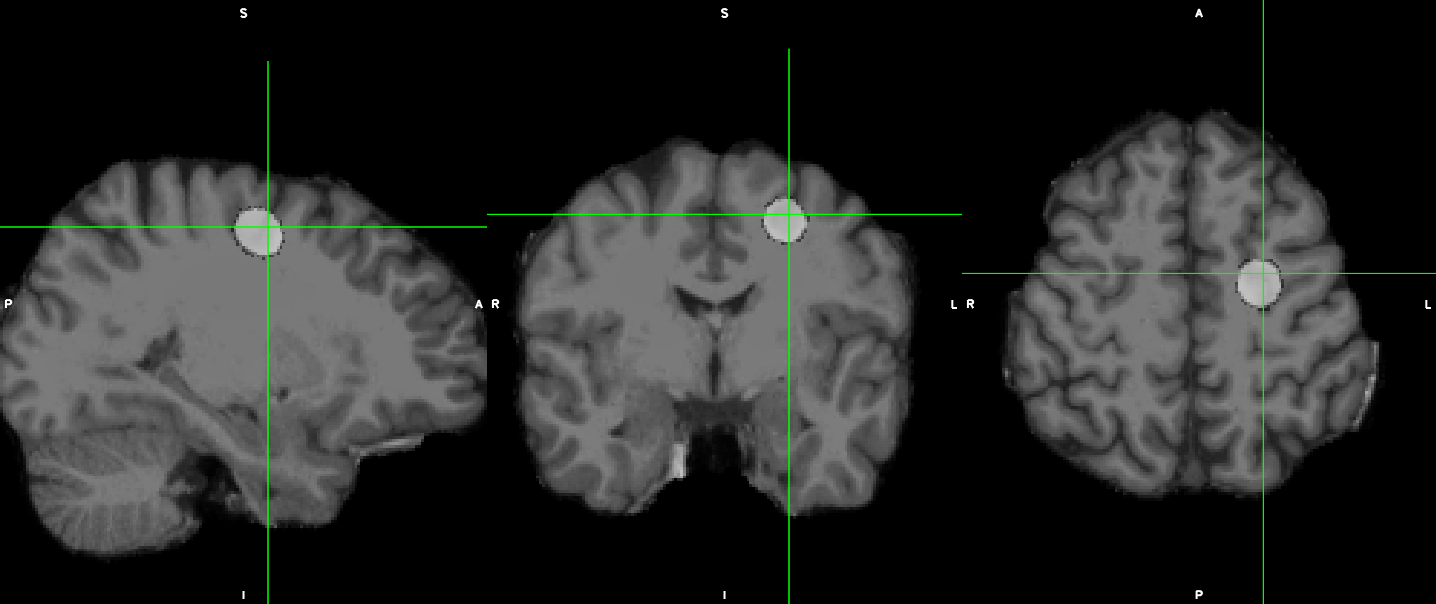}
  \caption{ Ellipse mask (right).}
  \label{fig:ellips} 
  \end{subfigure}
  \caption{FCD region annotations.}
\end{figure}

 It worth to mention that unlikely to previous works we have weak annotation of FCD subjects. The labeled subjects are annotated with 2D bounding boxes per each view (sagittal, coronal, axial) by professional radiologist (see fig. \ref{fig:ex_of_vis}). The input MR image is first preprocessed to correct the bias field,  align with a standard atlas (MNI152 1mm template \cite{mazziotta2001probabilistic}), and strip the non-brain tissues (FSL \cite{jenkinson2012fsl} and FreeSurfer \cite{reuter2012within} software). Also intensities were normalized with Histogram Standardization and Z-normalization using TorchIO library \cite{perez-garcia_torchio_2020}. To obtain 3D ground truth FCD regions, for each subject we fitted  parallelepiped into 2D boxes.

As a baseline model, we use the method proposed in \cite{wang2020automated}. It consists of 4 major steps: 1) preprocessing, 2) brain patch extraction, 3) deep learning classification, and 4) postprocessing. We suggest referring to the original article for further details. 

 Our main goal is to create an assisting tool for professional radiologists. As a key part of the algorithm is patch extraction, it seems natural to assess models by evaluating probabilities assigned to patches. So we introduce \emph{Top-k score} metric for model evaluation. At the inference stage, we take $k$ top patches by probabilities and estimate whether at least one of them intersects with the ground-truth FCD region. In that case, we mark it as successful detection. Then we average it across all validation subjects.

\subsection{Improvements}

The baseline model yielded a top-k score of 0.2. We made some iterative improvements to boost its performance. Careful ablation study is given in table \ref{result_table} (configurations a-g).

Data was labeled with 2D boxes on each of the axial, coronal, and sagittal slices. It is not correct to put a 3D rectangular box onto them, as, firstly, large non-FCD regions will be marked as FCD. Secondly, rectangularity is lost at the image alignment step.  To overcome this issue, we inscribe a 3D ellipse to use it as a mask (see fig. \ref{fig:ellips}). 

Furthermore, we refused to use hard labeling (i.e., any non-zero intersection leads to '1' labeling) for this task and, instead, use soft labeling. Ellipse mask still captures some non-FCD regions. Hence it is fair to give labels according to how big the overlap is. We calculated the number of overlapping voxels with an FCD mask for each patch, divided it by the maximum number for a given subject. We want labels for big-overlap patches to be close to 1, and labels for small-overlap patches to be around 0.5. This can be achieved by taking the fifth root of the acquired number.


Analyzing the patch extraction procedure, we noticed that size 16x32 is actually too small to cover all brain regions. Some FCD areas were just missed. So we increased the patch size from 16x32 to 24x40.

\begin{figure}[ht!]
  \centering
  \begin{subfigure}{\linewidth}
   \includegraphics[width=\linewidth]{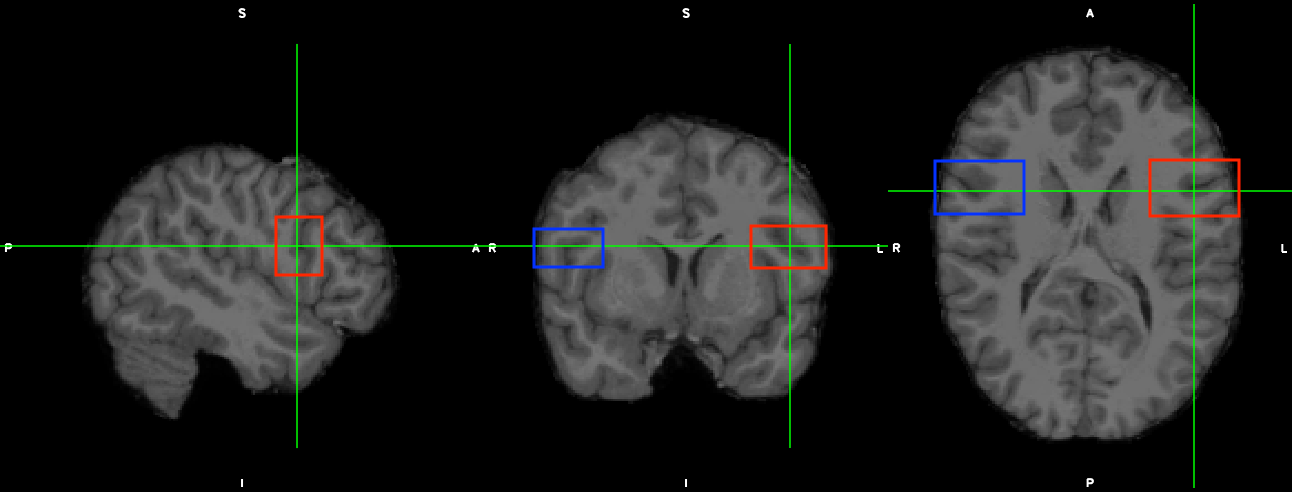}
  \caption{Example stack of patches taken from multi dimensions. Red - original patches. Blue - their 'mirrored' versions.}
  \label{fig:extra_views}     
  \end{subfigure}
  \begin{subfigure}{\linewidth}
        \includegraphics[width=\linewidth]{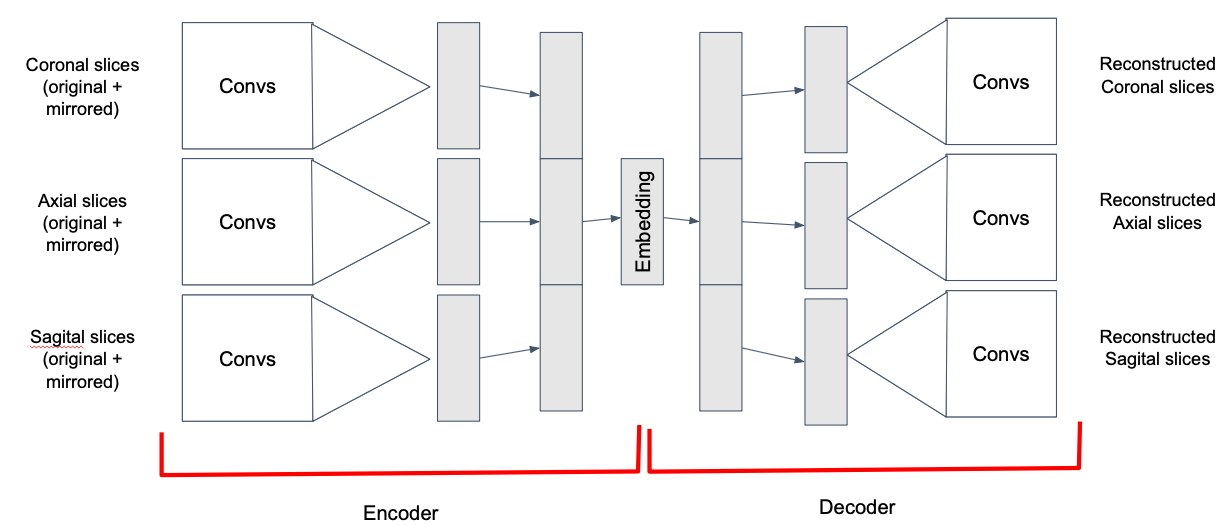}
  \caption{Autoencoder with '3 heads - 3 tails' architecture}
  \label{fig:ae}
  \end{subfigure}
  \caption{Multi-dimension patches' processing scheme.}
\end{figure}

To further boost performance, we asked ourselves why not to use all dimensions, taking slices from coronal and sagittal slices. For each axial slice, one should consider the axial slice itself, coronal slice crossing with it at the center, and sagittal slice crossing with it at the center, as well as their 'mirrored' neighbors (see fig. \ref{fig:extra_views}). Now each input tensor is of shape 6x24x40.

It worth to mention that 8 out of 15 subjects have FCD located in the temporal gyrus, while 7 out of 15 subjects have FCD located in the non-temporal region. Temporal and non-temporal FCD regions are very different from their nature. CNN model tends to overfit to temporal zone, which results in model yielding high probabilities for temporal patches of non-temporal subjects.  To overcome this issue, we separately train models for temporal and non-temporal FCD subjects. We will refer to it as \emph{ensembling specific-to-localization models}.

We also pretrain autoencoder to take advantage of 15 unlabeled FCD subjects of the same modality as the 15 labeled ones. It is incorrect to concatenate all slices into multi-channel image, so we use autoencoder with '3 heads - 3 tails' architecture (see \ref{fig:ae}). Each slice pair is processed by its own convolution, concatenating results and applying one more linear operation afterward. The decoder is a mirrored version of the encoder.

\section{Results}
\begin{figure}[!ht]
\centering
\begin{subfigure}{.2\linewidth}
    \includegraphics[width=\linewidth]{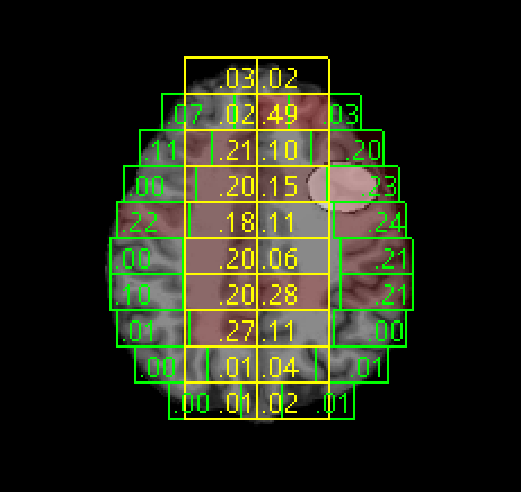}
    \caption{}\label{fig:image1}
\end{subfigure}
\begin{subfigure}{.2\linewidth}
    \includegraphics[width=\linewidth]{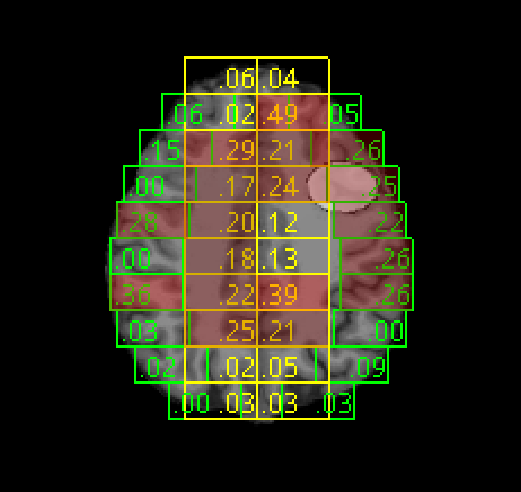}
    \caption{}\label{fig:image12}
\end{subfigure}
\begin{subfigure}{.2\linewidth}
    \includegraphics[width=\linewidth]{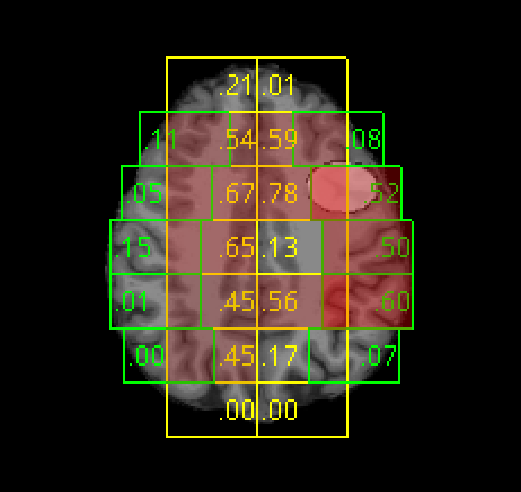}
    \caption{}\label{fig:image13}
\end{subfigure}
\begin{subfigure}{.2\linewidth}
    \includegraphics[width=\linewidth]{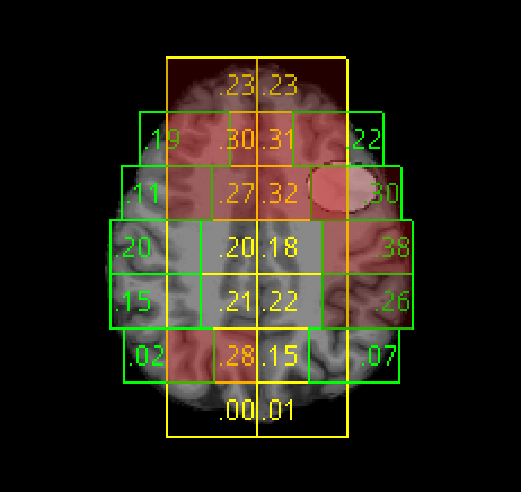}
    \caption{}\label{fig:image3}
\end{subfigure} 
\begin{subfigure}{.2\linewidth}
    \includegraphics[width=\linewidth]{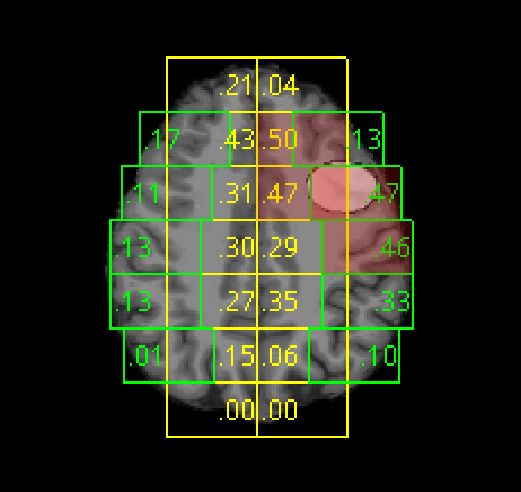}
    \caption{}\label{fig:image1}
\end{subfigure}
\begin{subfigure}{.2\linewidth}
    \includegraphics[width=\linewidth]{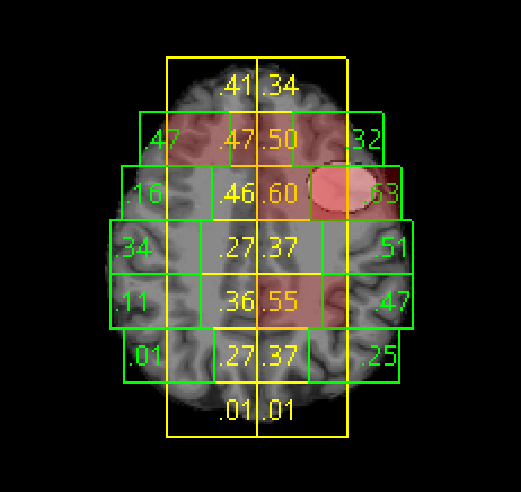}
    \caption{}\label{fig:config_f}
\end{subfigure}
\begin{subfigure}{.2\linewidth}
    \includegraphics[width=\linewidth]{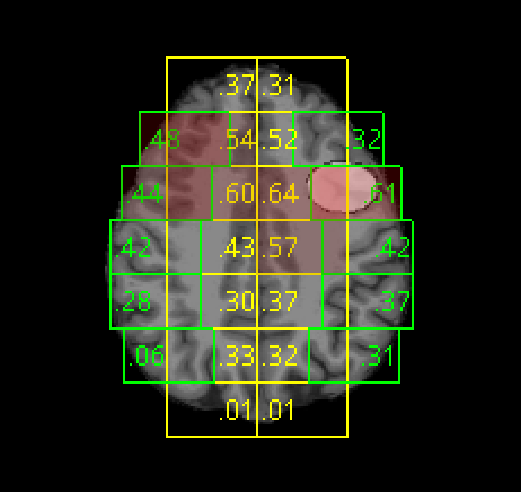}
    \caption{}\label{fig:image13}
\end{subfigure}
\begin{subfigure}{.2\linewidth}
    \includegraphics[width=\linewidth]{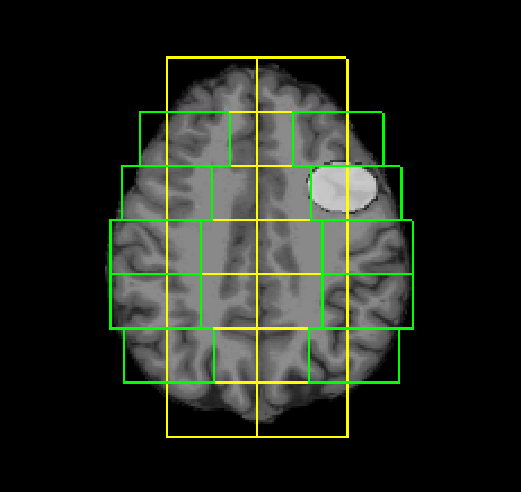}
    \caption{Original slice}\label{fig:image3}
\end{subfigure} 
\caption{Example of model performance on the same slice for different configurations (a)-(g). (h) - original slice without models' predictions. Green - side patches, yellow - middle patches, light ellipse - FCD region. For each patch probability yielded by model is shown. 
}
\label{fig:viz_comparison}
\end{figure}
\begin{table}[!ht]
\caption{Top-20 score ablation study  \label{result_table}}
\begin{tabularx}{\textwidth} { 
  | >{\raggedright\arraybackslash}X 
  | >{\raggedleft\arraybackslash}X  | }
  
 \hline
 Model & Top-20 score \\
 \hline
 (a) baseline model  & 0.200  \\
\hline
 (b) \quad + \quad soft patch labeling  & 0.266  \\
\hline
 (c) \quad + \quad increased patch size  & 0.400  \\
\hline
 (d) \quad + \quad autoencoder pretraining  & 0.533 \\
\hline
 (e) \quad + \quad specific-to-localization models  & \textbf{0.733}  \\
\hline
 (f) \quad + \quad coronal slices  & 0.667  \\
\hline
 (g) \quad + \quad sagittal slices  & \textbf{0.733} \\
\hline
\end{tabularx}
\end{table}
\begin{figure}[!ht]
\centering
\begin{subfigure}{.2\linewidth}
    \includegraphics[width=\linewidth]{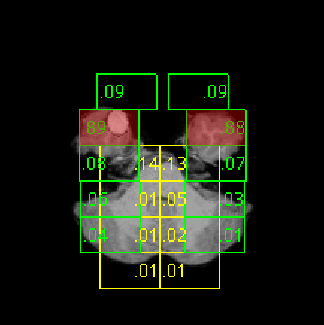}
    \caption{}\label{fig:image1}
\end{subfigure}
\begin{subfigure}{.2\linewidth}
    \includegraphics[width=\linewidth]{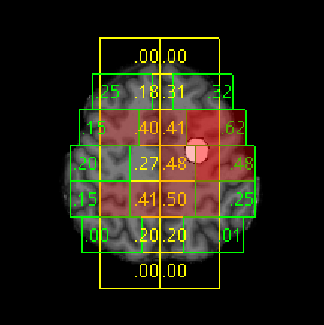}
    \caption{}\label{fig:subj1}
\end{subfigure}
\begin{subfigure}{.2\linewidth}
    \includegraphics[width=\linewidth]{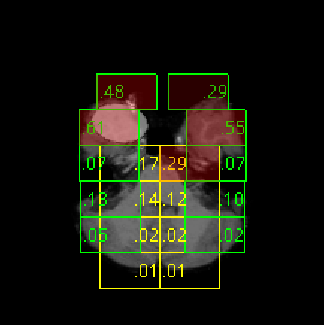}
    \caption{}\label{fig:image13}
\end{subfigure}
\begin{subfigure}{.2\linewidth}
    \includegraphics[width=\linewidth]{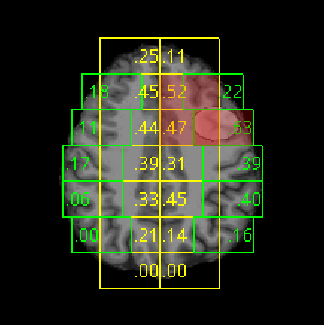}
    \caption{}\label{fig:image3}
\end{subfigure} 
\begin{subfigure}{.2\linewidth}
    \includegraphics[width=\linewidth]{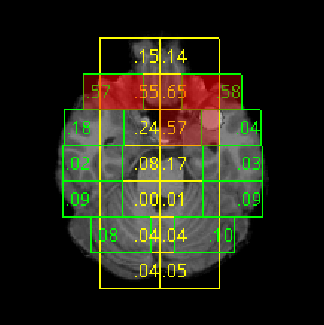}
    \caption{}\label{fig:image1}
\end{subfigure}
\begin{subfigure}{.2\linewidth}
    \includegraphics[width=\linewidth]{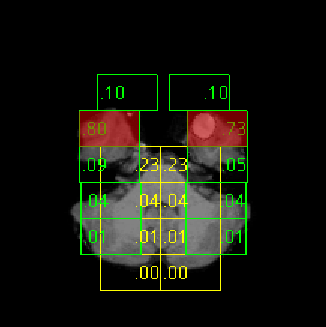}
    \caption{}\label{fig:qweqw}
\end{subfigure}
\begin{subfigure}{.2\linewidth}
    \includegraphics[width=\linewidth]{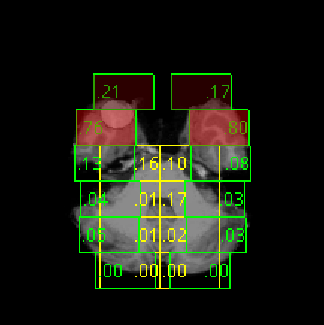}
    \caption{}\label{fig:subj11}
\end{subfigure}
\begin{subfigure}{.2\linewidth}
    \includegraphics[width=\linewidth]{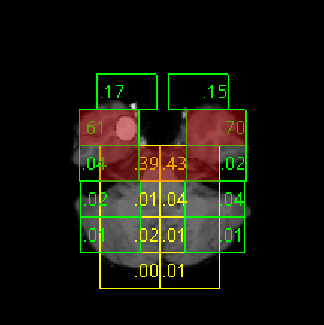}
    \label{fig:image3}
\end{subfigure} 
\caption{Example of performance of model with configuration  (e)  on 8 FCD subjects. }
\label{fig:conf_e}
\end{figure}

In table \ref{result_table}, it is shown that the largest boost in model performance was given by ensembling specific-to-localization models. Using coronal or sagittal slices did not lead to better results, though without ensembling specific-to-localization models, it did. 

For each configuration visualisation is provided on figure \ref{fig:viz_comparison}. For configuration (a), one can notice that model tends to give higher probabilities for middle patches.   Configurations (a)-(d) provides relatively high probabilities for non-FCD patches (False Positives), but captures desired. Configuration (f) tends to saturate the probabilities of patches. However, it does give the highest probability to FCD patch (see \ref{fig:config_f}). We did not notice an improvement in visualization for configurations (f) and (g) compared to (e).

We provide several examples of configuration (e) model's performance (see fig \ref{fig:conf_e}). Generally, model tends to capture all FCD-located patches. However, model has problem with distinguishing FCD patches and their mirrored versions, especially for ones located in temporal gyrus. Some FCD regions are predicted with relatively small probability. It is caused either by small size of FCD regions (\ref{fig:subj1}) or by lack of characteristic features of FCD, e.g. increased thickness or assymetry (\ref{fig:subj11}).

The best top-20 score is 0.733, so FCD regions were successfully detected for 11 out of 15 subjects. We recommend using configuration (e), as it is harder to interpret results by models using slices from extra dimensions  (f, g).


\section{Conclusion}


In the current work we reproduced a state-of-the-art DL-based method for FCD detection and proposed new metric for FCD detection algorithm's assessment. We investigated several ways to improve current approach specifically to deal with weakly annotated FCD data. We significantly boosted the FCD-detection CNN models performance with the use of soft labeling, autoencoder pretraining and ensembling specific-to-localization models. 

\clearpage

\bibliography{main_paper} 
\bibliographystyle{splncs03}

\end{document}